# Janus: Automatic Ontology Builder from XSD files

Ivan Bedini
Orange Labs,
France
(+33) 2 31 75 95 41
ivan.bedini@orange-ftgroup.com

Benjamin Nguyen
PRiSM Laboratory,
University of Versailles, France
(+33) 1 39 25 40 49
benjamin.nguyen@prism.uvsq.fr

Georges Gardarin
PRiSM Laboratory,
University of Versailles, France
(+33) 1 39 25 40 54
georges.gardarin@prism.uvsq.fr

## ABSTRACT
The construction of a reference ontology for a large domain still remains an hard human task. The process is sometimes assisted by software tools that facilitate the information extraction from a textual corpus. Despite of the great use of XML Schema files on the internet and especially in the B2B domain, tools that offer a complete semantic analysis of XML schemas are really rare. In this paper we introduce Janus, a tool for automatically building a reference knowledge base starting from XML Schema files. Janus also provides different useful views to simplify B2B application integration.

## Categories and Subject Descriptors
D.3.3 [**Programming Languages**]: Language Constructs and Features – *Data models, Semantics, Dynamic storage management.*

## 1. INTRODUCTION
Over the past ten years, the Semantic Web wave has shown a new vision of ontology use for application integration systems. Researchers have produced several software tools for building ontologies (like Protégé [1]) and merging them two by two (like Mafra [2]) or producing alignments (like S-Match [3]). These solutions, as well as adopted ontology building methodologies, are mainly human or, as shown in [4], sometimes assisted by semi-automatic software tools.

Limitations to their adoption for integration of internet and enterprise applications, among others reasons, are: (i) the lack of tools capable of extracting and acquiring information from a collection of XML files (the "de-facto" format for applications information exchange definition); (ii) the complexity of aligning and merging two or more knowledge sources, a task excessively consuming of computational time; (iii) the difficulty of validation based on background knowledge hard to produce and maintain.

Janus, the tool that we have developed, implements a new approach to ontology generation capable of providing a solution to the limitations described above. It is able to automatically generate and maintain, (in a meta-model OWL compatible), a collective memory resource that facilitates the discovery of alignments for matching concepts in a given domain. The knowledge source is automatically acquired from an extensible set of XSD files. The aim of this short paper is to introduce Janus, how it works and to show some different views (produced by the tool) of knowledge automatically acquired.

## 2. JANUS ARCHITECTURE
Our tool implements an adaptation of several techniques originating from text mining and information retrieval/extraction fields, applied to XML files (that we call **XML Mining**). XML Mining is used to pre-process simple and compound statements defining XML tags, such as XSD elements and XSD complex types. It includes clustering methods based on a Galois Lattice to quickly discovery similarities between names and structures extracted from XML schemas.

Figure 1 shows the overall architecture of Janus.

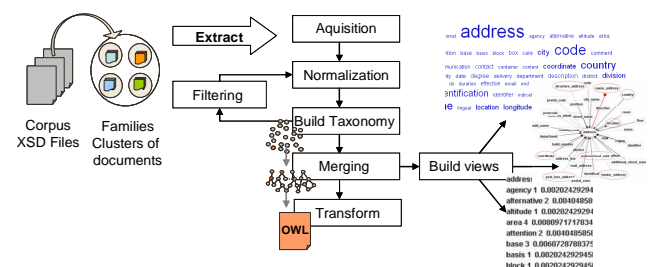

**Figure 1 - Janus overall architecture**

The algorithm generating a high level representation of XML Schema information sources is composed of three main steps. The first step is the **Extraction** task represented by the *Extract* arrow and *Acquisition* rectangle in Figure 1. It provides the knowledge needed to generate the ontology (concepts, properties and relationships). Implemented techniques for knowledge acquisition are a combination of different types, such as: *NLP* (Natural Language Process) for morphological and lexical analysis, *association mining* for calculating term frequencies and association rules, *semantics* for finding synonymy, and *clustering* for grouping semantic and structural similar concepts.

The second step is **Analysis** represented by the *Normalisation*, *Build Taxonomy* and *Filtering* blocks in Figure 1. This step focuses on the matching of retrieved information and/or alignment of two or more existing families of concepts issued from different sources. This step requires techniques already used in the first stage, as syntax and semantic measures, to establish the best similarities; it also requires an analysis of concept structures to determine hierarchical relationships and identify common properties.

The last step is **Generation**, represented by the *Merging*, *Transform* and *Build View* blocks in Figure 1. This step looks for concepts with evident affinities (e.g., concept fully included into another) to merge them. It transforms the meta-model used by the tool into a semantic network that can be described in RDFS or OWL. The tool can derive from the network useful views provided to users.

Users can also step into the process to parameterize thresholds for refining results.

## 3. FUNCTIONALITIES and VIEWS

The tool currently offers three visualization methods to view the acquired knowledge. The **word view** shows the list of terms composing the "ontology", and tag clouds (both in Figure 2). The **table view** gives detailed information about each concept (like frequencies, family attendance, source instances and direct relationships). The **graph view** diplays the semantic network (Figure 3). Other views, "Concepts Social Network", are under development.

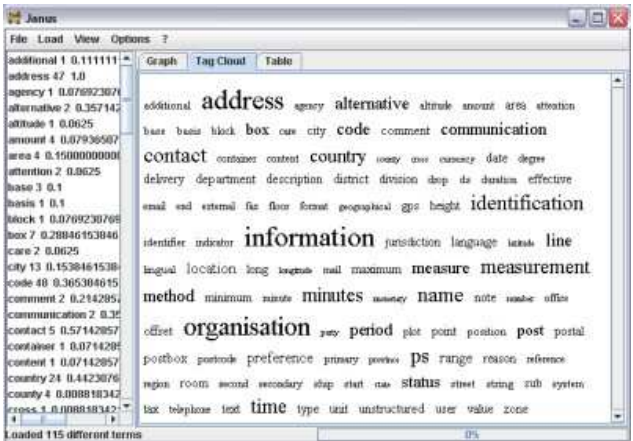

**Figure 2 - Janus GUI – List and Tag Cloud Overview**

The graph view can show the whole graph or only the part related to selected concepts with different layouts (hierarchical, tree, …).

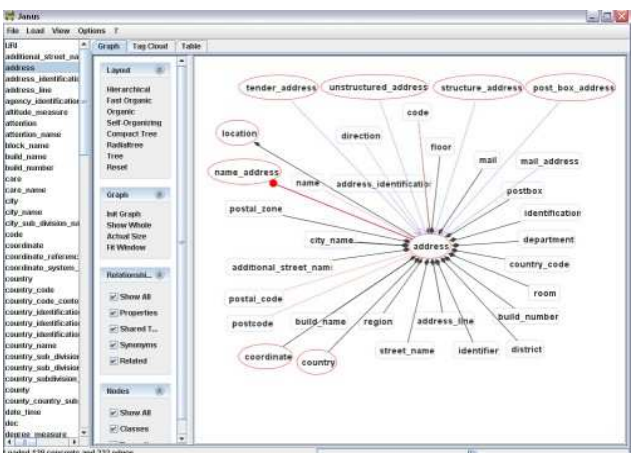

**Figure 3 - Janus GUI Overview**

Acquired relationships, and thus visualised, are of different types: *propertyOf*, *synonym*, *shared terms* (compound tags with common terms like *address* for *tender_address* and *post_box_address*) and *relatedTo* (mainly merged concepts or other of type "owl:sameAs").

The graph view also provides the possibility to choose the relationships to highlight, as well as concepts classes and/or properties (Figure 4). This feature permits to analyse in details some parts of the ontology; it is useful when the ontology is too large to be browsed with the global view.

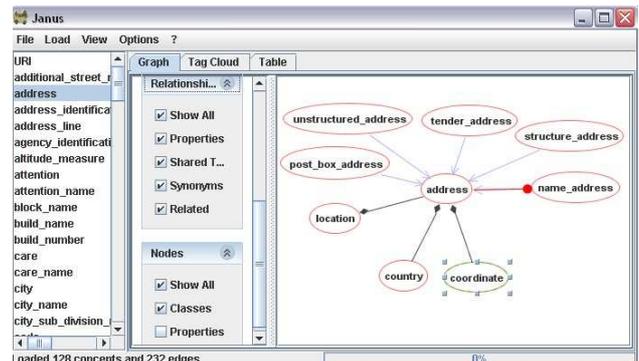

**Figure 4 - Janus GUI Overview**

The tool also offers the possibility to parameterize thresholds for alignment and merging operations.

## 4. CONCLUSION

The automatic acquisition of knowledge from XML Schema files shows encouraging results, as well as the implemented meta-model (a semantic model detailed in [5]) that looks robust enough to continue research works in this direction. Before developing a complete and automatic tool, several problems remain to be resolved. But nevertheless, the ability to acquire semantic concepts directly from the "interchange language" (XML) without the *a priori* definition of application ontologies, is a new approach to B2B integration. Janus provides a first implementation of this approach.

We propose to demonstrate the first results of this tool applied to the analysis of 23 existing B2B standards.

## 5. REFERENCES

[1] N. F. Noy, R. W. Fergerson, & M. A. Musen. The knowledge model of Protege-2000: Combining interoperability and flexibility. EKAW 2000, France.

[2] Maedche, A., Motik, B., Silva, N., and Volz, R.. MAFRA - Mapping Distributed Ontologies in the Semantic Web. In Proc. EKAW 2002.

[3] Fausto Giunchiglia, Mikalai Yatskevich, and Pavel Shvaiko. Semantic matching: Algorithms and implementation. Journal on Data Semantics, IX, 2007.

[4] Ivan Bedini and Benjamin Nguyen. Automatic Ontology Generation: State of the Art. PRiSM Laboratory Technical Report, University of Versailles, 2007.

[5] Ivan Bedini, Benjamin Nguyen and Georges Gardarin. Building Reference Ontologies from B2B XML Schema Files. PRiSM Laboratory Technical Report, University of Versailles, 2007.